\documentclass[twocolumn]{aastex631}

\usepackage{booktabs}
\usepackage{amsmath}

\begin{document}

\title{Investigating the AGN variability timescale - black hole mass relationship with Gaia, SDSS and ZTF}

\author[0000-0002-2620-6483]{Adrien Hélias}
\affiliation{Department of Physics \& Astronomy,\\ University of Western Ontario, 1151 Richmond St,\\
London, N6A 3K7, Canada}

\author[0000-0001-6217-8101]{Sarah C. Gallagher}
\affiliation{Department of Physics \& Astronomy,\\ University of Western Ontario, 1151 Richmond St,\\
London, N6A 3K7, Canada}
\affiliation{Institute for Earth and Space Exploration,\\ University of Western Ontario, 1151 Richmond St,\\
London, N6A 3K7, Canada}

\author[0000-0003-2767-0090]{Pauline Barmby}
\affiliation{Department of Physics \& Astronomy,\\ University of Western Ontario, 1151 Richmond St,\\
London, N6A 3K7, Canada}
\affiliation{Institute for Earth and Space Exploration,\\ University of Western Ontario, 1151 Richmond St,\\
London, N6A 3K7, Canada}

\begin{abstract}

Active galactic nuclei (AGNs) exhibit variability in their luminosities with timescales that correlate with the mass of the black hole at the centre of the AGN. Presently, the empirical correlation lacks sufficient precision to confidently convert these timescales into black hole masses, especially at the low-mass end. To find more AGNs with timescale measurements, we study a very large catalog of AGNs from the Gaia Data Release 3 called GLEAN (Gaia variabLE AgN; $872~228$ objects). We identify GLEAN objects with optical spectra from the Sloan Digital Sky Survey DR17 and light curves from the Zwicky Transient Facility (ZTF) DR21.  After fitting the light curves with a damped random walk model, we find that the GLEAN light curves have insufficient sampling to extract reliable amplitude and timescale measurements outside the range of 50--100 days. On the other hand, well-sampled ZTF light curves allow more accurate estimations of these parameters. The fractional variability amplitude is an effective, model-independent metric for measuring variability amplitude, but only when derived from high-quality light curves. We provide a catalog of 127 GLEAN AGNs with spectroscopic virial black hole masses, and variability amplitudes and timescales.  Though we do not find any low-mass black holes in this AGN sample, we confirm a relationship between the damped random walk timescale and the black hole mass that is consistent with previous studies.

\end{abstract}


\section{Introduction} \label{sec:intro}

Active galactic nuclei (AGN) are growing supermassive black holes located at the centres of galaxies \citep{Lynden-Bell1969,SilkRees1998}. They emit a considerable amount of light and may generate powerful winds and jets of matter \citep[e.g.,][]{Faucher2012,Blandford2019}. Most of the optical and ultraviolet light from active galaxies does not come from the stars; it comes from the accretion disk of the supermassive black hole (SMBH) \citep{SunMalkan1989,LaorNetzer1989,Hubeny2001}. The entire emission of the AGN spans the electromagnetic spectrum from radio to X-ray wavelengths \citep[e.g.,][]{Padovani2017}. All active AGNs exhibit some variability in luminosity regardless of their mass and physical size \citep{Ulrich1997}. This variability arises from changes in the accretion disk continuum emission; subsequently, the fluxes of the broad emission lines in AGN spectra respond after a time lag \citep{Peterson1982}. This time delay is due primarily to the light travel time from the accretion disk to the broad-line region (BLR) \citep{Peterson2001}. A part of this time delay is due to the time for the BLR gas to respond, but this is expected to be minor compared to the light travel time delay.\\

Optical variability is a relatively recent method used to find AGNs that is effective even for X-ray-faint AGNs. The processes driving the optical continuum variability occur on timescales of hours to years depending on the physical size of the AGN accretion disk \citep{Peterson1982}. Stochastic models applied to AGN light curves such as Continuous Auto-Regressive Moving Average (CARMA) models \citep{Kelly2014}, which include the popular damped random walk (DRW) model, have proven to be useful tools to characterize variability timescales \citep{Kelly2009, Kozlowski2010,Macleod2010,Kelly2011,Baldassare2018,Moreno2019}. It has also been shown that the variability can deviate from these models \citep{Mushotzky2011,Zu2013,Kasliwal2015,Guo2017,Stone2022,Su2024}.\\ 

CARMA models are not physically based but are still useful empirical mathematical tools for extracting more information from the data. Furthermore, the parameters have been shown to correlate with the AGN properties. For example, several researchers have found an empirical relationship between the variability timescale and the black hole mass of the AGN \citep{Kelly2009,Macleod2010,Kozlowski2016,Guo2017,Suberlak2021,Burke2021,Arevalo2024,Su2024}. \citet{Burke2021} in particular (B21 hereafter), found a positive correlation between the rest-frame DRW timescale and the black hole mass (see Figs.~1 and 3 in B21). B21 kept 67 AGNs in their final sample, with high-quality light curves and a mix of reverberation-mapped and single-epoch spectroscopic black hole masses. Most of their AGNs possess black hole masses between a few 10$^7$ and 10$^9~M_{\odot}$. However, the slope of the correlation is primarily determined by the small batch of points located between 10$^4$ and 10$^7~M_{\odot}$. \citet{Wang2023} added 79 data points of AGNs located in dwarf galaxies to the B21 sample. These new points lie between 10$^6$ and 10$^{7.5}~M_{\odot}$ in black hole mass, and are consistent with the best-fit line of B21. Nevertheless, AGNs with a central black hole mass in the $10^2-10^6~M_{\odot}$ range are poorly observed and are needed to further constrain the relationship at low masses. Being able to confidently infer a black hole mass from a DRW timescale would be extremely useful, particularly in cases where AGN have weak or undetectable broad lines. This is especially the case for low-mass, low-luminosity AGNs. More data are required to confirm or contradict the full shape of the relationship.\\

To better constrain the rest-frame DRW timescale - black hole mass correlation of B21 and \citet{Wang2023}, we studied a very large catalog of AGNs called GLEAN (Gaia variabLE AgN; \citealt{GLEAN2023}). GLEAN is the first Gaia catalog of variable AGNs, coming from the Gaia Data Release 3 (Gaia DR3; \citealt{GaiaDR3_2023}). It contains astrometric, photometric and rough spectroscopic information on $872~228$ extragalactic objects, which were selected solely on Gaia variability. The GLEAN catalog includes light curves for each object, one per filter, with irregular sampling, taken over the course of 3 years. Most importantly, \cite{GLEAN2023} estimated that the purity of the sample, which is the number of genuine AGNs divided by the number of objects in the catalog, is at least 95\%, with conservative assumptions. This is an immense, all-sky collection of AGNs selected on variability; likely some of them are new AGN identifications. Previous studies have used Gaia DR3 to perform varstrometry -- a technique to trace the apparent movement of the centroid of the total flux from a source when it exhibits variability in its light curve -- to identify unresolved off-nucleus or lensed quasars on sub-kpc scales \citep{Chen2023, LambertSecrest2024}. Others have used Gaia DR3 as a secondary catalog for finding dual AGNs with the multipeak method \citep{Wu2024}.\\

In this work, we explore and characterize the GLEAN sample by cross-matching it with the Sloan Digital Sky Survey (SDSS) \citep{SDSS2000} and the Zwicky Transient Facility (ZTF) \citep{ZTF2019}, as well as studying the correlation between the central black hole mass and the variability timescale with a subset of GLEAN AGNs. In Section~\ref{sec:glean}, we describe the GLEAN sample and we cross-match GLEAN with SDSS to analyze the SDSS classification of the GLEAN AGNs and their properties. In Section~\ref{sec:ztf}, we detail the stochastic modelling performed to obtain timescales and amplitudes from light curves, and we cross-match GLEAN with ZTF to obtain well-sampled light curves for our AGNs. We calculate virial black hole masses from SDSS optical spectra, and we compare our own fit to the sample data with the B21 correlation. We discuss the results in Section~\ref{sec:discussion}, and finally, we draw our conclusions in Section~\ref{sec:conclusion}.

\section{Characterization of GLEAN}
\label{sec:glean}

\begin{figure*}[ht!]
\centering
\includegraphics[width=0.8\textwidth]{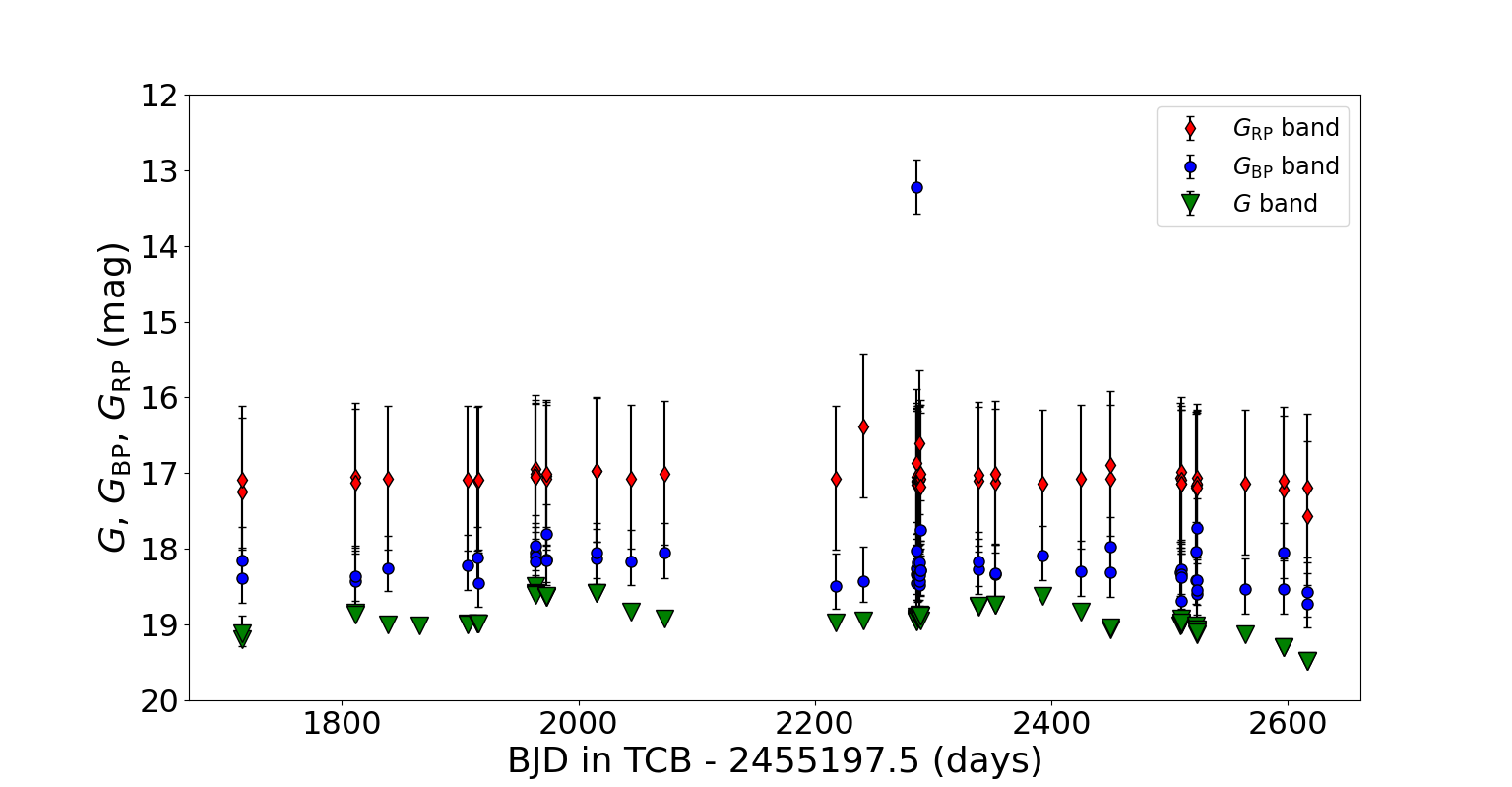}
\caption{Light curve of the GLEAN object with Source ID 1268847039908643840 in Gaia DR3. The green dots represent the $G$-band, the blue dots the $G_{\mathrm{BP}}$-band and the red dots the $G_{\mathrm{RP}}$-band. BJD in TCB is the Barycentric Julian Date in Barycentric Coordinate Time, or the number of days elapsed since January 1st, 2010. The isolated blue point is representative of the outliers that appear in many Gaia light curves and is likely instrumental rather than physical. The sampling is uneven and is different for each object.}
\label{fig:GLEAN light curve}
\end{figure*}

The Gaia space observatory uses 3 photometric filters: $G$, $G_{\mathrm{BP}}$ and $G_{\mathrm{RP}}$. The $G_{\mathrm{BP}}$-band ($\sim$ 330--680 nm) and the $G_{\mathrm{RP}}$-band ($\sim$ 640--1050 nm) are subsets of the larger $G$-band, which spans $\sim$ 330--1050 nm \citep{GaiaMission2016}. The GLEAN catalog includes light curves for each object, one per filter, with irregular sampling, taken over the course of $\sim$ 3 years ($\sim 1000$ days). Figure~\ref{fig:GLEAN light curve} shows a typical light curve for one of the AGNs in the GLEAN catalog. The selection process began with a pool of 34 million potential sources within Gaia DR3, pinpointed as possible AGN candidates by 11 distinct classifiers that assessed their variability through supervised machine learning.  \citet{GLEAN2023} relied solely on data from Gaia and used the Gaia Celestial Reference Frame 3 (Gaia-CRF3). They identified sources without significant proper motion and parallax measurements as extragalactic. Then, they set a criterion of 20 for the minimum number of photometric data points in the $G$-band, along with essential variability metrics such as the index of the structure function and the \citet{ButlerBloom2011} criteria. The median uncertainties for RA/Dec positions are 0.01--0.02 mas for $G < 15$, 0.05 mas at $G = 17$, 0.4 mas at $G = 20$, and 1 mas at $G = 21$ mag \citep{GaiaDR3_2023}. Positional cross-matches are thus quite accurate. The GLEAN sample is therefore an interesting parent sample for AGN studies, with an extremely large sample size and coverage of the whole sky.\\

The GLEAN light curves measure the optical variability in luminosity of the targeted objects between 2014 and 2017. The fractional variability amplitude, $f_G$ (in the $G$-band), quantifies the amplitude of variability as calculated with the following formula:
\begin{equation}
f_G = \frac{\sqrt{\mathrm{MAD}(F)^2 - <\sigma_F^2>}}{\mathrm{Median}(F)}~,
\label{eq:f_G}
\end{equation}
\noindent where $F$ is the flux, MAD is the median absolute deviation, and $\sigma_F$ is the flux uncertainty \citep{GLEAN2023}. This is not the classic definition of the fractional variability amplitude \citep{Vaughan2003}, as the mean was replaced by the median in the denominator and the variance was replaced by the MAD. These changes were made to limit the effects of outlier data points in the light curves.  Outliers are extremely common and would otherwise change the value of $f_G$ dramatically. The quantity, $f_G$, along with other parameters such as the structure function index, its scatter, and the \citet{ButlerBloom2011} metrics for quasar variability, were used to make cuts in the Gaia DR3 database and only keep the variable AGN candidates.\\

\begin{figure}[ht!]
\centering
\includegraphics[width=0.5\textwidth]{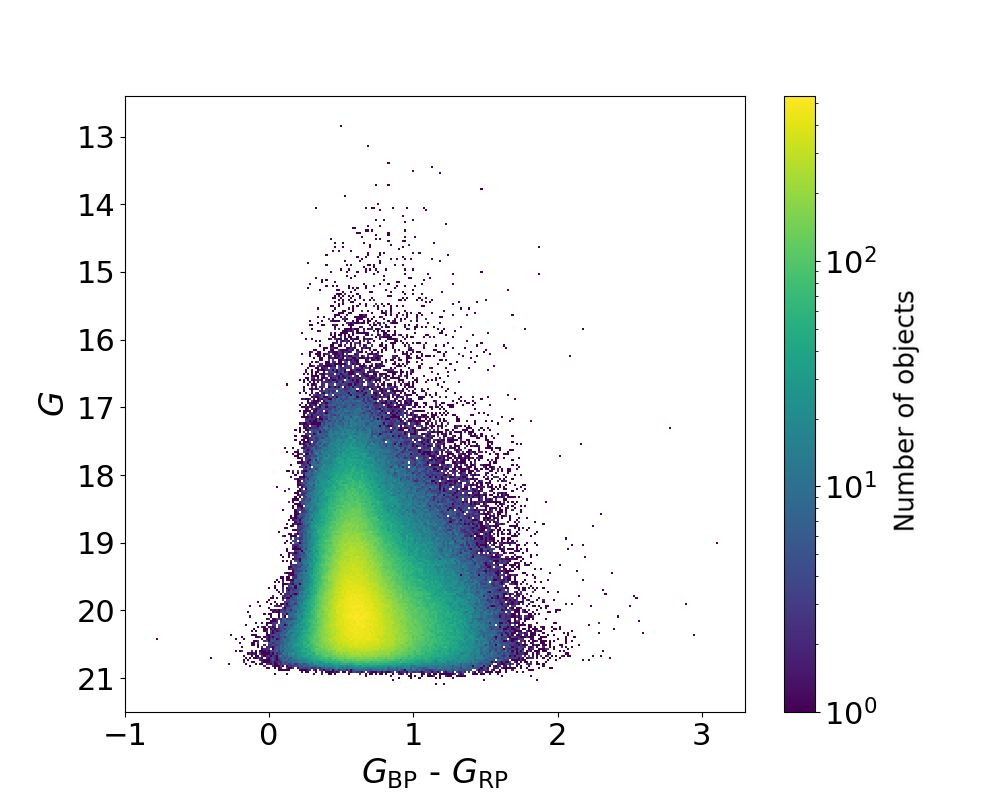}
\caption{Color-magnitude diagram of the $872~228$ GLEAN variable AGNs. The faintest magnitude in GLEAN is $G \sim 21$. Fainter objects tend to be redder.}
\label{fig:G mag}
\end{figure}

\begin{figure}[ht!]
\centering
\includegraphics[width=0.5\textwidth]{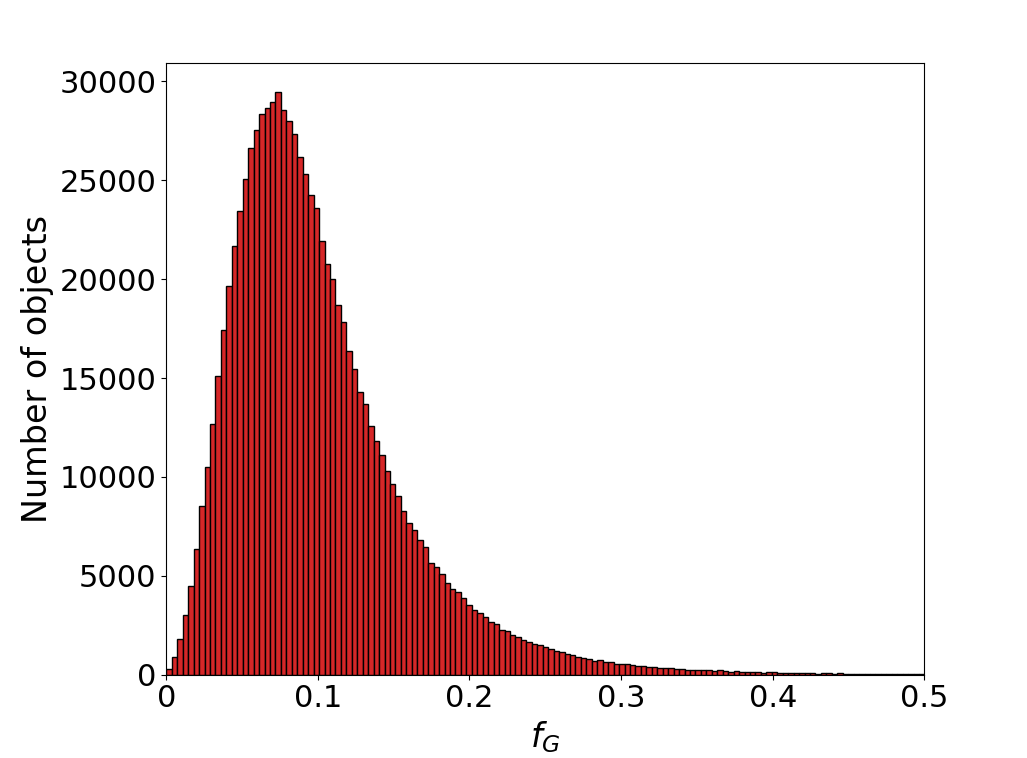}
\caption{Histogram of fractional variability amplitude $f_G$ in the $G$-band of GLEAN variable AGNs. Most objects have $f_G \sim 0.08$, and 93.9\% of the distribution has 0 $\lesssim f_G \lesssim$ 0.2.}
\label{fig:Histo frac g}
\end{figure}

\begin{figure}[ht!]
\centering
\includegraphics[width=0.5\textwidth]{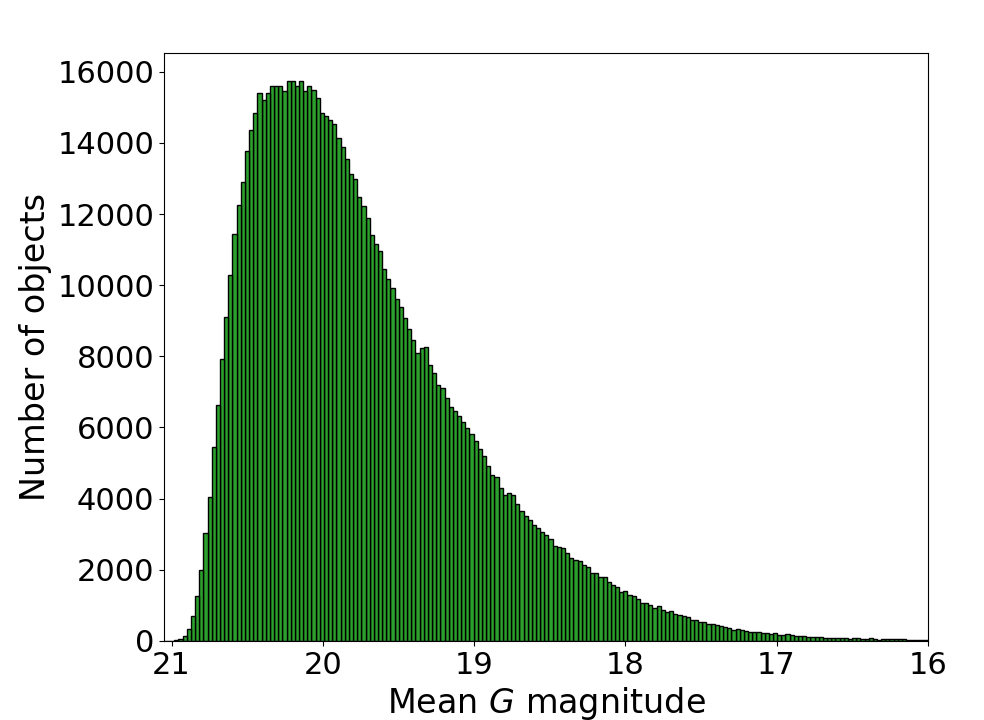}
\caption{Histogram of mean $G$ magnitude of GLEAN variable AGNs. Most objects have $G \sim 20.2$, and 84.2\% of the distribution has 19 $\lesssim G \lesssim$ 21.}
\label{fig:Histo mean mag}
\end{figure}

\begin{figure}[ht!]
\centering
\includegraphics[width=0.5\textwidth]{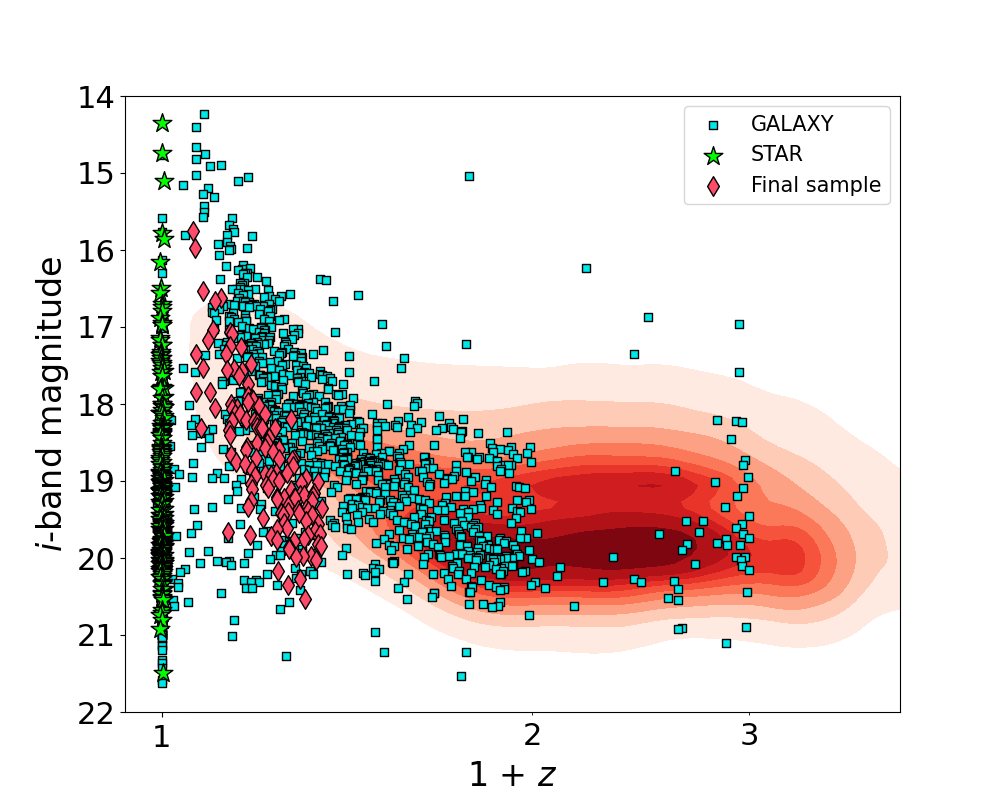}
\caption{SDSS $i$-band model magnitude versus redshift (1 + $z$) for the $203~915$ GLEAN-SDSS objects. The colors represent different classes: the red density contours for QSOs ($201~631$ objects), the blue squares for GALAXYs ($2050$ objects) and the green stars for STARs ($234$ objects). The 127 AGNs of the final sample are shown in pink salmon diamonds.}
\label{fig:ibandz}
\end{figure}

\begin{figure}[ht!]
\centering
\includegraphics[width=0.5\textwidth]{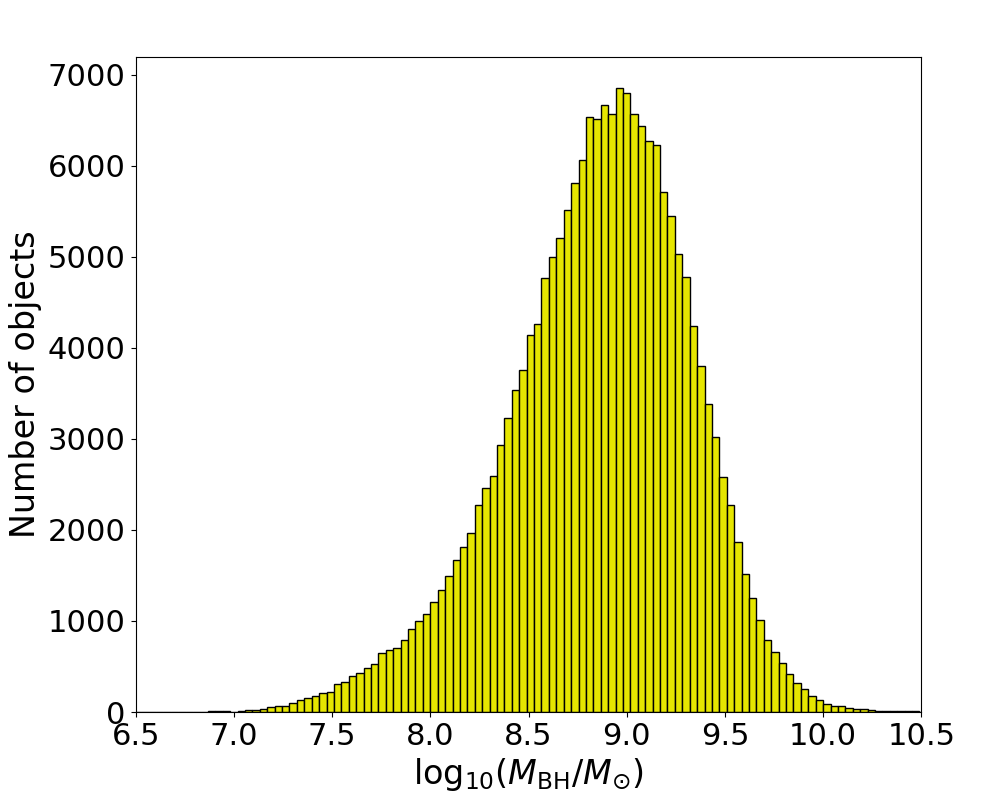}
\caption{Black hole mass histogram of the $195~943$ GLEAN AGNs that are present in the \citet{WuShen2022} SDSS DR16 quasar catalog. The maximum is attained at $M_{\mathrm{BH}} = 10^9 M_{\odot}$ with no AGNs with $M_{\mathrm{BH}}<10^{6.7} M_{\odot}$. The distribution is clearly skewed towards the high-mass end of the supermassive black hole population.}
\label{fig:bhmasses}
\end{figure}

Figures~\ref{fig:G mag}--\ref{fig:bhmasses} characterize the GLEAN catalog. Figure~\ref{fig:G mag} presents a color-magnitude plot of the GLEAN sample. The faintest magnitude observed in GLEAN is approximately $G \sim 21$. As seen in Figure~2, fainter sources typically exhibit more pronounced emission in the red portion of the optical spectrum when compared to the blue part while brighter sources are bluer. Figure~\ref{fig:Histo frac g} is a histogram of the fractional variability amplitude $f_G$ for the GLEAN catalog. The distribution peaks at $f_G \sim 0.08$, and 93.9\% of the sources have $0 \lesssim f_G \lesssim 0.2$. Objects with a large variability amplitude are therefore rare in this sample, but because GLEAN is so large, there are still hundreds of objects with high values of $f_G$ (e.g., $f_G \gtrsim 0.4$). Likewise, Figure~\ref{fig:Histo mean mag}, a histogram of the mean $G$ magnitude, exhibits the same shape. The distribution of the median $G$ magnitude is consistent with Figure~\ref{fig:Histo mean mag}. As expected, bright AGNs are rare compared to faint ones. The peak of the distribution is $G \sim 20.2$, and 84.2\% of the sources have 19 $\lesssim G \lesssim$ 21.\\

We then cross-match GLEAN with the Sloan Digital Sky Survey to check the classification. Some objects labeled as variability-selected AGNs in GLEAN may not be detected as AGNs in SDSS. We use the spectroscopic data of the 17th Data Release of the Sloan Digital Sky Survey for this analysis (SDSS DR17; \citealt{SDSSDR17}). SDSS has the following system of classes and subclasses for each spectrum:
\begin{itemize}
\item GALAXY: The spectrum is best fit with a galaxy template.
\begin{itemize}
\item STARFORMING: The galaxy has detectable emission lines that are consistent with star formation: $\mathrm{log}_{10}(\mathrm{OIII/H}\beta) < 0.7 - 1.2~\mathrm{log}_{10}(\mathrm{NII/H}\alpha) - 0.4$.
\item STARBURST: The galaxy is star-forming but has an equivalent width of H$\alpha$ greater than 50 $\mathrm{\AA}$.
\item AGN: The galaxy has detectable emission lines that are consistent with a Seyfert or a low-ionization nuclear emission-line region (LINER): $\mathrm{log}_{10}(\mathrm{OIII/H}\beta) > 0.7 - 1.2~\mathrm{log}_{10}(\mathrm{NII/H}\alpha) - 0.4$.
\end{itemize}
\item QSO: The spectrum is best fit with a QSO template.
\item STAR: The spectrum is best fit with a stellar template, and has star subclasses (F5, OB, K7, M3, CV...)
\end{itemize}

The templates are built from a rest-frame principal-component analysis (PCA) of training samples of known redshift\footnote{https://www.sdss4.org/dr17/algorithms/redshifts/}. Finally, there is also a BROADLINE subclass that can be added to any GALAXY or QSO with emission lines detected at the 10-sigma level with a Gaussian feature width $\sigma > 200~\mathrm{km~s^{-1}}$ at the 5-sigma level.\\

We evaluate the SDSS classes for the $203~915$ objects that compose the cross-match between GLEAN and SDSS DR17. The large majority of them (98.88\%) are defined as QSOs, as expected given the high purity of GLEAN. Small but non-negligible fractions of this sample are GALAXY (1\%) and STAR (0.12\%). Some of these could be contaminants such as star-forming galaxies, stars, white dwarfs, and some others have bad or noisy spectra, but there could also be wrongly classified AGNs. $188~813$ objects have a BROADLINE subclass (92.6\%), and 277 objects have an AGN subclass (0.14\%). Further investigation on these objects not recognized as AGNs is required to unveil their true nature and will be part of future work.\\

Characterizing the GLEAN sample further, we plot the $i$-band magnitude versus redshift distribution of the GLEAN-SDSS objects coded by spectral class in Figure~\ref{fig:ibandz}. Most of the GLEAN-SDSS objects have $1 + z < 3$, with a peak at $1 + z \sim 2.5$. QSOs cover most of the parameter space up until $1 + z \sim 4$, located between $17 < i < 21$, with a peak at $i \sim 20$. GALAXYs have a more complex distribution, with three local maxima: ($1 + z \sim 1$, $i \sim 19.7$), ($1 + z \sim 1.3$, $i \sim 18.4$) and ($1 + z \sim 2.9$, $i \sim 19.7$). These are likely a consequence of selection effects from the way SDSS galaxies are chosen for spectroscopy (e.g., luminous red galaxies have a fainter flux limit for spectroscopic follow-up \citep{Eisenstein2001}). Finally, STARs are grouped in a very thin stripe at $1 + z \sim 1$. If a spectrum has been wrongly classified using a very different template compared to its true nature, then the redshift measurement is also likely to be wrong (e.g., featureless spectra, which can be ``fitted'' by a variety of different templates for different redshifts).\\

\citet{WuShen2022} compiled a catalog of quasar properties from SDSS DR16 (DR16Q) including $750~414$ broad-line quasars. Of these, $195~943$ overlap with GLEAN (22.5\% of GLEAN AGNs), and we present their black hole mass distribution in Figure~\ref{fig:bhmasses}, using \citet{WuShen2022} measurements from the H$\beta$, \ion{Mg}{2} and \ion{C}{4} emission lines. The distribution is very smooth with a maximum at $M_{\mathrm{BH}} \approx 10^9 M_{\odot}$. This figure shows that the GLEAN AGNs in DR16Q are very clearly in the high-mass part of the supermassive black hole population. None of them have SMBH masses below $10^{6.7} M_{\odot}$. This implies that, while it will be very unlikely to add low-mass AGNs to the variability timescale - black hole mass relationship, we should have plenty of candidates to start uncovering the true shape of the relationship in the supermassive regime.\\

\section{Variability timescales and black hole masses} \label{sec:ztf}

The ZTF light curves and SDSS spectra can be used to obtain variability timescales and black hole mass measurements, respectively. We start with stochastic modelling of GLEAN light curves fit with a damped random walk (DRW) model defined by the following differential equation:
\begin{equation}
\frac{dx(t)}{dt} + \alpha_1 x(t) = \beta_0 \epsilon(t)~,
\end{equation}
where $x(t)$ is the state of the system, $\alpha_1$ is the auto-regressive coefficient, $\beta_0$ is the moving average coefficient and $\epsilon(t)$ is a source of noise. The DRW model is the CARMA(1,0) model. The ($p, q$) coefficients of a CARMA($p, q$) model provide the order of the auto-regressive process and the moving average process, respectively. The coefficient of the highest order derivative $d^px(t)/dt^p$, $\alpha_0$, is typically set to 1. In the case of the DRW model, $\alpha_1$ is linked to a relaxation timescale $\tau_{\mathrm{relax}} = \tau_{\mathrm{DRW}} = 1/\alpha_1$, and $\beta_0 = \beta_{\mathrm{DRW}}$ is the amplitude of variability.\\

We perform DRW modelling using the Python package \textit{taufit} (\href{https://github.com/burke86/taufit}{github.com/burke86/taufit}) to fit AGN light curves, with the Markov Chain Monte-Carlo (MCMC) technique. It gives three output parameters: $\beta_{\mathrm{DRW}}$, $\tau_{\mathrm{DRW}}$, and $\sigma_n$, the excess white noise amplitude. Therefore, \textit{taufit} determines the white noise present in the light curve. B21 performed simulations of light curves comparable to the cadence and baseline of ZTF light curves, and they identified four selection criteria to ensure reliability in the $\tau_{\mathrm{DRW}}$ and $\beta_{\mathrm{DRW}}$ measurements from DRW modelling:
\begin{enumerate}
    \item Baseline condition: The baseline ($t_{\rm max}-t_{\rm min}$) is at least ten times longer than $\tau_{\mathrm{DRW}}$.
    \item Cadence condition: The average cadence ($\Delta t$) is shorter than $\tau_{\mathrm{DRW}}$.
    \item Signal-to-noise ratio (SNR) condition: $\beta_{\mathrm{DRW}}$ is larger than the average flux uncertainties $\overline{dy}$ ($\beta_{\mathrm{DRW}} > \sqrt{\overline{dy}^2 + \sigma_n^2}$).
    \item Auto-correlation condition: The auto-correlation function (ACF) of the light curve is different from the ACF of a white noise signal with $3\sigma$ confidence.
\end{enumerate}
To compare our results with B21 more easily, we use the same four selection criteria. Since \textit{taufit} uses the package \textit{emcee}, we use 32 walkers with 500 burn-in samples and 2000 production samples.\\

\begin{figure*}[ht!]
\centering
\includegraphics[width=\textwidth]{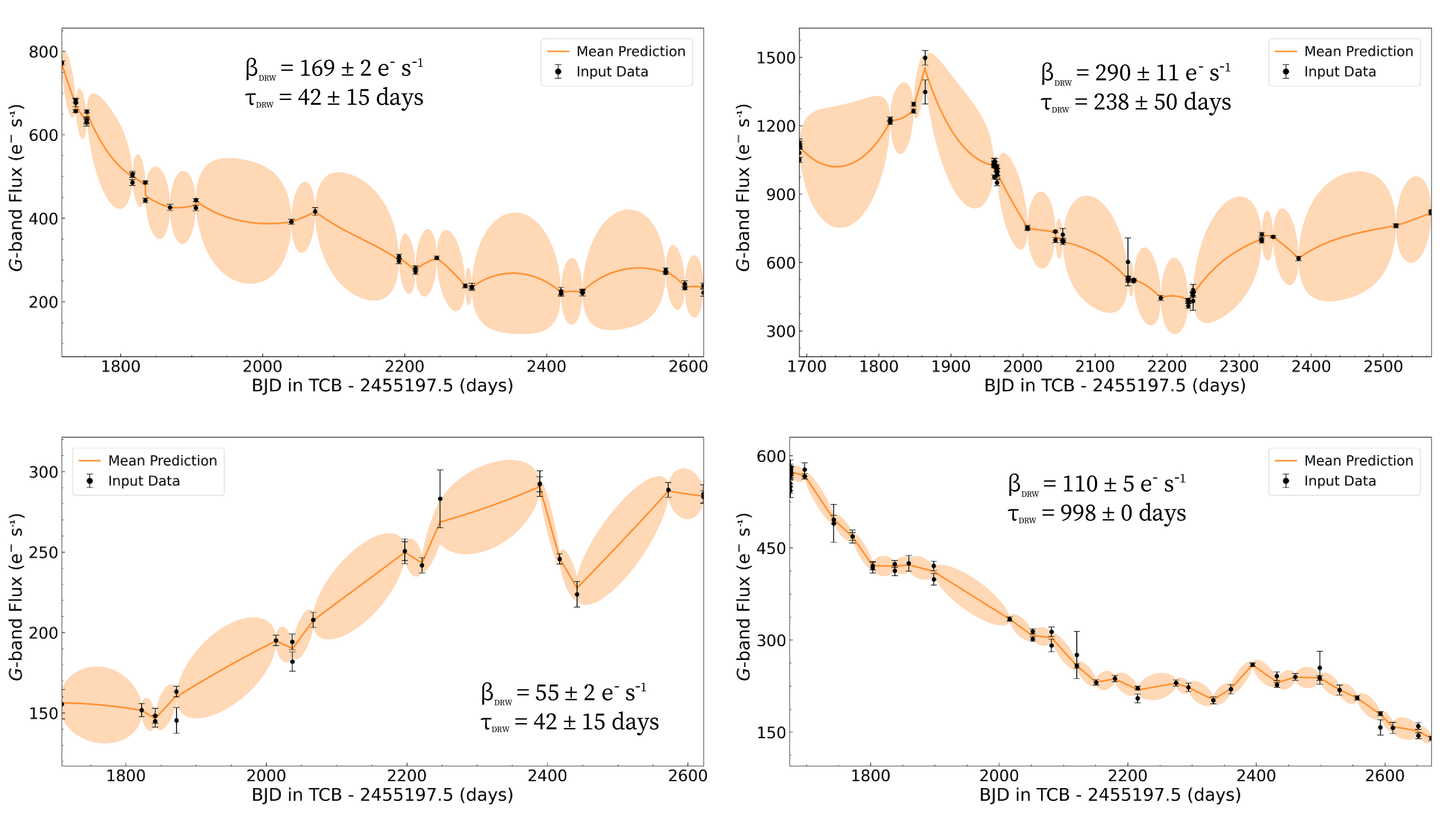}
\caption{Damped random walk modelling of four GLEAN light curves in the $G$-band (Source IDs of the objects from left to right and top to bottom: 1199811036972579456, 3705764960573610112, 6334030370128440576, 4726105218503224448). The black dots with uncertainties are the light curve data points. The orange line is the mean prediction from the damped random walk fitting, with the orange area representing the 1$\sigma$ dispersion. The retrieved amplitude $\beta_{\mathrm{DRW}}$ and timescale $\tau_{\mathrm{DRW}}$ are displayed on each panel. The global shape of the light curves is captured, but the lack of shorter cadence sampling tends to generate timescales which are systematically overestimated.}
\label{fig:example drw fits}
\end{figure*}

Figure~\ref{fig:example drw fits} presents fits selected at random of four GLEAN light curves with the DRW model. The uncertainties of the fit can be relatively large during periods without observations. The general shape of the light curve is captured, but the small variations through short periods of time are not accounted for. If the AGN has a true variability timescale longer than the light curve itself, or if the light curve is not sampled well enough to show a few stochastic variations, the code will tend to converge to a long timescale, between 800--1000 days. Multiple works have shown how the DRW timescale can be biased by the baseline of the light curve \citep{Kozlowski2017,Suberlak2021,Burke2021,Hu2024,Zhou2024}. The bottom right figure in Figure~\ref{fig:example drw fits} demonstrates this issue very well. The code could not identify a pattern in the light curve, resulting in a very long estimated timescale of 998 days --- almost the total length of the curve --- with an unrealistic uncertainty of $\pm$ 0 days. The lack of shorter-cadence sampling in the GLEAN light curves means that timescales tend to be systematically overestimated.\\

After testing the $872~228$ GLEAN light curves for DRW modelling, we find that they are not well-sampled enough to accurately derive timescales outside of the range 50--100 days. Their main flaw is the limited number of data points with respect to the light curve baseline (length). A 3-year ($\sim$1000 days) light curve for all $872~228$ AGNs of GLEAN is remarkable, but there is a necessary trade-off in sampling rate. Most GLEAN light curves have between 20 and 60 data points per band. If the cadence were regular, this would be equivalent to having only one data point every 17 to 50 days. This makes it impossible to obtain reliable small timescales below 50 days, according to the cadence condition. Also, the baseline of the light curves prevents calculating trust-worthy long timescales over $\sim$100 days, according to the baseline condition. Therefore, we conclude that GLEAN light curves are not ideal for stochastic modelling.\\

To see how the damped random walk modelling performs on higher-quality light curves for GLEAN AGNs, we cross-match GLEAN with the Data Release 21 of ZTF (ZTF21; \citealt{ZTF2019}). ZTF is an automatic time-domain survey which scans the Northern sky in the optical with the Palomar 48-inch telescope. The primary science targets of ZTF are asteroids, transients and variable sources \citep{Bellm2019}. ZTF uses the $ugriz$ filters and the light curves can be in the $g$-band, the $r$-band and the $i$-band, though not all objects have data in these three filters. For this paper, we only use the $g$-band light curves as they typically have the largest number of data points, and AGNs are more variable at shorter wavelengths. We use a 1" radius to cross-match GLEAN with ZTF. Out of the $872~228$ GLEAN AGNs, $626~204$ cross-matches were found with ZTF21. Not all $626~204$ objects in the GLEAN-ZTF cross-match have many data points in their light curves. Some AGNs have $\sim 2000$ epochs while others have zero, but the distribution is skewed towards low epoch numbers. We discard observations with the bad quality flag 32768, because these measurements are likely to be unusable (e.g., affected by clouds and/or moonlight).\\

In addition to fitting the ZTF light curves of GLEAN AGNs, we calculate their virial (single-epoch) black hole masses using optical spectroscopy. We do this because not all AGNs in our sample have black hole mass measurements in DR16Q: 676 825 GLEAN objects are not in it, which represents 77.5\% of GLEAN. Typically, the broad component of the H$\beta$ line for $z \lesssim 0.7$ galaxies \citep{Kaspi2000} is used for black hole mass estimation. However, H$\beta$ is not always ideal because it may be weak, especially compared to the H$\alpha$ line, which is often at least three times stronger \citep{Greene2005}. Furthermore, H$\beta$ is often blended with the \ion{Fe}{2} complex and [\ion{O}{3}], and the flux from the host galaxy can contaminate the broad H$\beta$ component. Instead, estimating the black hole mass with the H$\alpha$ line has been a successful method in recent decades, particularly for AGN spectra with notable stellar continuum emission \citep{Greene2004,Greene2005,Greene2007,Reines2013,Chilingarian2018}; this is the method we are using for GLEAN AGNs. There may be some stellar absorption lines around H$\alpha$, but the broad H$\alpha$ luminosities we work with (10$^{41}$--$10^{43}$ erg s$^{-1}$) are high enough such that the absorption lines would only affect the narrow component of H$\alpha$. Therefore, we do not need to subtract the starlight for our black hole mass measurements, because we only use the broad component of the H$\alpha$ line, which only comes from AGNs.\\

To obtain optical spectra for GLEAN AGNs, we cross-match our sample with the 17th Data Release of the Sloan Digital Sky Survey (SDSS DR17; \citealt{SDSSDR17}) using RA/Dec sky coordinates within a search radius of 1$^{\prime\prime}$.$200~989$ of the GLEAN-ZTF cross-matches have a spectrum in SDSS DR17. To get the best results, we only keep 493 GLEAN AGNs with at least 1000 epochs in their ZTF $g$-band light curve ($n_{\rm obs} \ge 1000$). We are only retrieving spectra flagged as PRIMARY, which is designed to choose the best available unique set of spectra for a given location on the sky. We correct each spectrum for the redshift, using the spectroscopic redshift values from SDSS. We also obtain the $E(B-V)$ values to account for Galactic extinction using the recent Corrected Schlegel–Finkbeiner–Davis (CSFD) dust map from \citet{Chiang2023}. We focus on the region from rest-frame 642 to 671 nm, which includes only the H$\alpha$ line at 656.3 nm and the two [\ion{N}{2}] lines at 648.2 nm and 661.1 nm. The [\ion{S}{2}] doublet is not taken into account because it is outside of our wavelength range. A model with four Gaussian components is applied to the fit: two Gaussians for the [\ion{N}{2}] lines, one for the narrow H$\alpha$ component, and one for the broad H$\alpha$ component. The Gaussian fits to the [\ion{N}{2}] doublet have the same width values (with a very narrow margin). [\ion{N}{2}] and H$\alpha$ have a fixed separation with each other. The continuum is fitted by a linear function of the form $f(\lambda) = a\lambda + b$ with parameters ($a$, $b$). We use the software Sherpa\footnote{\href{https://sherpa.readthedocs.io/en/4.16.0}{https://sherpa.readthedocs.io/en/4.16.0}} to perform the fitting. If $z \ge$ 0.356 for a given AGN, we discard it from the spectral fitting because the H$\alpha$ line is not present in the SDSS optical spectrum or is too close to the long wavelength limit. We also discard the objects with flat (i.e., no evidence of emission lines) or noisy spectra, and the objects without insufficient signal-to-noise to fit the model. Specifically, if the continuum fit (between 642 and 671 nm) is above $30\times10^{-17}~\mathrm{erg~s^{-1}~cm^{-2}~\AA^{-1}}$, the spectrum is either not from an AGN or polluted by a nearby star, and we discard the spectrum. We also remove spectra with fewer than 10 data points above five times the standard deviation of the continuum flux density between 642 and 671 nm. In total, we discard 165 out of 493 objects for all the reasons explained above. Finally, when we apply the four aforementioned conditions of \citet{Burke2021} to keep a given light curve in our sample after DRW modelling, we end up with 127 AGNs. The cadence and baseline conditions are the ones that filter out the most objects, followed closely by the SNR condition. The 127 spectra have been visually inspected, and all sources of the final sample are showing a broad line in the H$\alpha$ emission line. They are bona-fide broad line AGNs.\\

To calculate the black hole mass, $M_{\mathrm{BH}}$, we extract the area and the full-width half-maximum (FWHM) of the broad-line component from the broadline Gaussian model. We use the conservative formula from \citet{Chilingarian2018} based on the work of \citet{Reines2013}. It is conservative because the radius-luminosity relationship on which the $M_{\mathrm{BH}}$ formula relies uses careful assumptions on distance estimates, impacted by large uncertainties from the peculiar velocities of galaxies and surface brightness \citep{Bentz2013}. The $M_{\mathrm{BH}}$ formula is:
\begin{equation}\label{eq:BH_mass}
\begin{split}
& M_{\mathrm{BH}} = \\
& 3.72 \times 10^6 \left(\frac{L_{\mathrm{H\alpha}}}{10^{42}~\mathrm{erg~s^{-1}}}\right)^{0.47}\left(\frac{\mathrm{FWHM}_{\mathrm{H\alpha}}}{10^{3}~\mathrm{km~s^{-1}}}\right)^{2.06} M_{\odot}~,
\end{split}
\end{equation}
\noindent where FWHM$_{\mathrm{H\alpha}}$ is the FWHM of the broad H$\alpha$ component, corrected for the FWHM of the SDSS instrument by the following formula:
\begin{equation}
\mathrm{FWHM}_{\mathrm{H\alpha}} = \sqrt{\mathrm{FWHM}_{\mathrm{observed}}^2 - \mathrm{FWHM}_{\mathrm{instrument}}^2}~,
\end{equation}
\noindent and $L_{\mathrm{H\alpha}}$ is the luminosity of the broad H$\alpha$ line:
\begin{equation}
\begin{split}
L_{\mathrm{H\alpha}} & = 4\pi D_{\mathrm{L}}^2 F_{\mathrm{H\alpha}} \\
& = 4\pi D_{\mathrm{L}}^2  a\sigma\sqrt{2\pi} \\
& = 4\pi D_{\mathrm{L}}^2 a\frac{\mathrm{FWHM}_{\mathrm{H\alpha}}}{2\sqrt{2\mathrm{ln}(2)}}\sqrt{2\pi}~,
\end{split}
\end{equation}
\noindent where $D_{\mathrm{L}}$ is the luminosity distance in cm, $a$ is the integrated flux under the broad component in $\mathrm{erg~s^{-1}~cm^{-2}~\AA^{-1}}$ and $\sigma$ is the standard deviation of the broad component in $\mathrm{\AA}$. $\mathrm{FWHM}_{\mathrm{instrument}}$, or $\Delta\lambda$, is the stated spectral resolution of the SDSS and Baryon Oscillation Spectroscopic Survey (BOSS) spectrographs\footnote{Both the SDSS and BOSS spectrographs were used to obtain optical spectra for the SDSS programs, and they have a resolution $R$ = $\lambda/\Delta\lambda$ = 1500 at 380 nm and $R$ = 2500 at 900 nm. We determine the value of $R$ for each spectrum by calculating the average value at the observed position of H$\alpha$, based on the $R$ values at 380 and 900 nm (\href{https://www.sdss4.org/dr17/spectro/spectro\_basics}{https://www.sdss4.org/dr17/spectro/spectro\_basics}).}. The statistical and systematic uncertainties of Equation~\ref{eq:BH_mass} have been estimated as 0.3 dex when compared to other methods of measuring the black hole mass such as stellar dynamics and reverberation mapping \citep{Xiao2011,Dong2012}. Therefore, this value will be our uncertainty estimate for virial black hole masses.\\

\begin{figure*}[ht!]
\centering
\includegraphics[width=\textwidth]{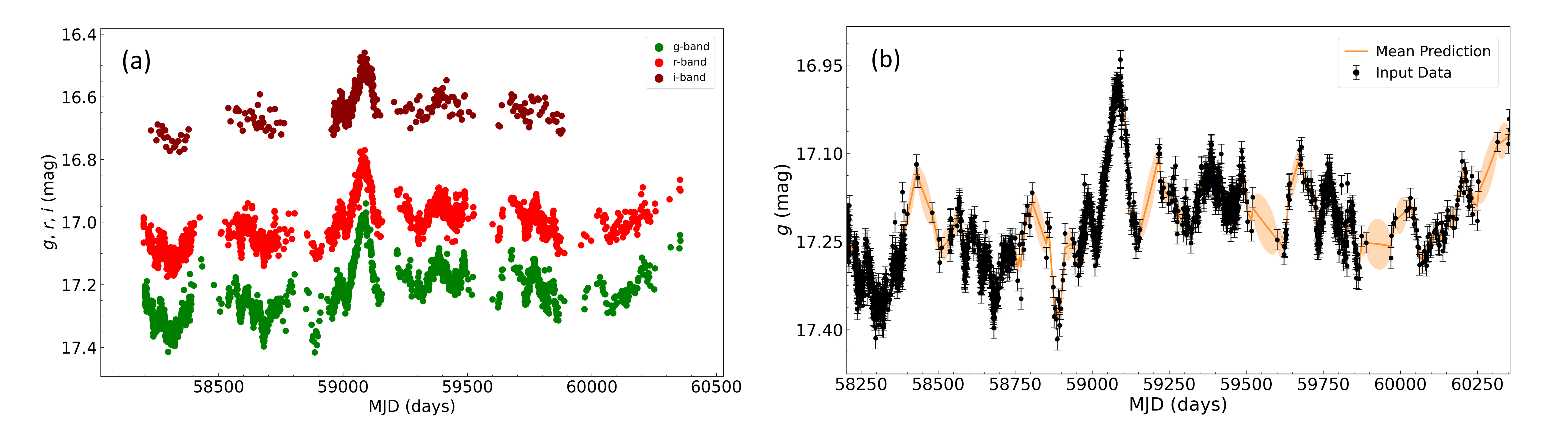}
\caption{(a) ZTF light curve of SDSS J170306.47+452557.2 containing data in the $g$-band, $r$-band and $i$-band, and (b) damped random walk modelling of the same ZTF light curve in the $g$-band. The black dots with uncertainties are the light-curve data points. The orange line is the mean prediction from the damped random walk fitting, with the orange area representing the 1$\sigma$ dispersion. MJD is the Modified Julian Date. ZTF light curves are well-sampled and suitable for a deep analysis with stochastic modelling.}
\label{fig:ZTF light curve}
\end{figure*}

Figure~\ref{fig:ZTF light curve}a presents a sample ZTF light curve with data in each band. This light curve is twice as long as GLEAN light curves ($\sim$ 6 years), with an irregular cadence ranging from a few hours to a month between observations. Figure~\ref{fig:ZTF light curve}b shows the DRW modelling of the same light curve, in the $g$-band. Because of the long baseline and the shorter average cadence between observations, the orange areas of uncertainty over the light curve time gaps are much more constrained than in the GLEAN light curves such as the one shown in Figure~\ref{fig:GLEAN light curve}. The damped random walk model fits the high-cadence ZTF light curves well.\\

\begin{figure*}[ht!]
\centering
\includegraphics[width=\textwidth]{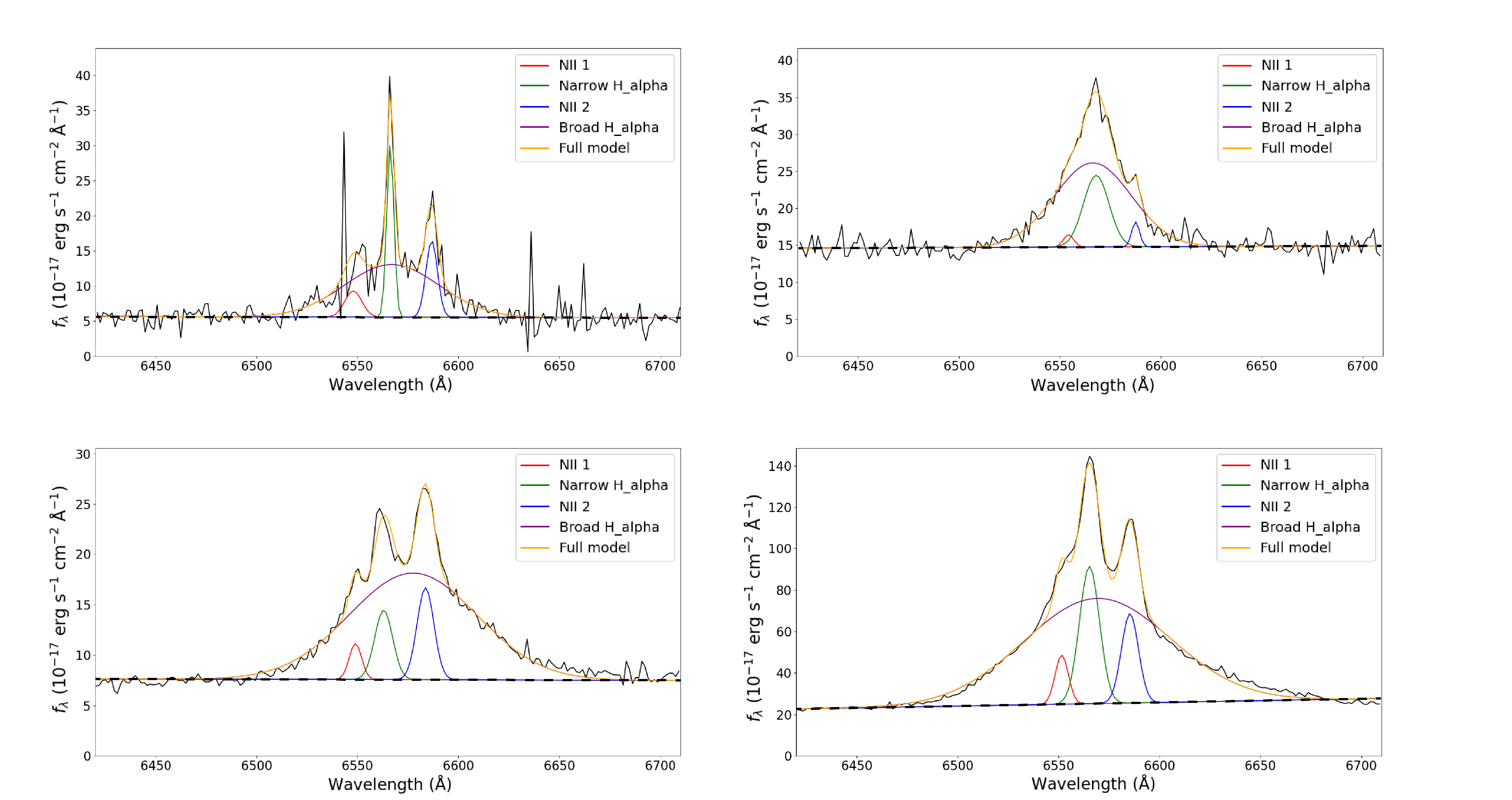}
\caption{Fitting of four SDSS spectra in our sample (SDSS designations from left to right and top to bottom: SDSS J144927.24+431644.1, SDSS J115451.10+233456.8, SDSS J122811.89+514622.8, SDSS J121822.72+385043.4) around the H$\alpha$ region. Wavelength is in the rest-frame.  The black solid line is the optical spectrum, the black dashed line is the best fit of the continuum, the red and blue curves are the [\ion{N}{2}] Gaussian components, the green curve is the narrow component of H$\alpha$, the purple curve is the broad component of H$\alpha$, and the yellow curve is the sum of the multi-component Gaussian model. All of the 127 GLEAN AGNs display broad lines.}
\label{fig:SDSS spectra}
\end{figure*}

\begin{deluxetable*}{rrrrrrrrrr}
\tablecaption{Final sample of 127 GLEAN AGNs for the variability timescale - black hole mass correlation.}
\tablehead{
\colhead{Name} & \colhead{$z$} & \colhead{$M_{\mathrm{BH}}$\tablenotemark{a}} & \colhead{$L_{\mathrm{H\alpha}}$} & \colhead{$L_{\mathrm{bol}}/L_{\mathrm{Edd}}$} & \colhead{$\mathrm{FWHM}_{\mathrm{H\alpha}}$} & \colhead{$\tau_{\mathrm{DRW}}$} & \colhead{$\beta_{\mathrm{DRW}}$} & \colhead{$\sigma_n$} & \colhead{$f_\mathrm{ZTF}$}\\
\colhead{} & \colhead{} & \colhead{($10^7 M_{\odot}$)} & \colhead{($10^{41} \mathrm{erg~s^{-1}}$)} & \colhead{} & \colhead{($\mathrm{km~s^{-1}}$)}& \colhead{(days)} & \colhead{(mag)} & \colhead{(mag)} & \colhead{}}
\startdata
SDSS J173038.28+550016.8 & 0.249 & 23.1 & 14.7 & 0.01 & 6796 & $66^{+32}_{-18}$ & $0.18\pm0.03$ & 0.12 & 0.11 \\
SDSS J172320.69+564349.1 & 0.269 & 5.6 & 16.3 & 0.05 & 3328 & $119^{+95}_{-39}$ & $0.13\pm0.03$ & 0.09 & 0.08 \\
SDSS J094118.22+582824.3 & 0.280 & 29.1 & 23.3 & 0.01 & 6844 & $3^{+1}_{-1}$ & $0.14\pm0.01$ & 0.14 & 0.09 \\
SDSS J082815.68+440611.2 & 0.338 & 7.9 & 21.7 & 0.05 & 3696 & $123^{+133}_{-45}$ & $0.13\pm0.04$ & 0.10 & 0.09 \\
SDSS J211646.34+110237.5 & 0.081 & 14.0 & 19.1 & 0.02 & 5020 & $177^{+233}_{-71}$ & $0.10\pm0.03$ & 0.04 & 0.08 \\
SDSS J122325.49+524600.8 & 0.345 & 14.7 & 55.0 & 0.05 & 4038 & $70^{+37}_{-19}$ & $0.10\pm0.02$ & 0.09 & 0.09 \\
SDSS J100543.59+521957.0 & 0.230 & 2.3 & 11.6 & 0.09 & 2326 & $119^{+162}_{-50}$ & $0.11\pm0.03$ & 0.10 & 0.04 \\
SDSS J100731.30+554919.3 & 0.173 & 3.8 & 7.0 & 0.04 & 3352 & $62^{+53}_{-27}$ & $0.11\pm0.02$ & 0.10 & 0.06 \\
SDSS J102339.66+523349.7 & 0.136 & 3.1 & 22.5 & 0.12 & 2312 & $53^{+26}_{-16}$ & $0.06\pm0.01$ & 0.03 & 0.05 \\
SDSS J115019.54+534724.2 & 0.061 & 3.7 & 4.8 & 0.03 & 3611 & $15^{+4}_{-3}$ & $0.07\pm0.01$ & 0.04 & 0.06 \\
SDSS J115538.75+534054.9 & 0.213 & 6.2 & 23.2 & 0.06 & 3242 & $136^{+139}_{-49}$ & $0.13\pm0.04$ & 0.06 & 0.08 \\
SDSS J120415.96+560258.1 & 0.090 & 2.9 & 7.0 & 0.05 & 2925 & $17^{+7}_{-4}$ & $0.06\pm0.01$ & 0.04 & 0.04 \\
SDSS J121720.33+544327.2 & 0.226 & 19.5 & 24.4 & 0.02 & 5576 & $139^{+149}_{-50}$ & $0.16\pm0.05$ & 0.08 & 0.14 \\
SDSS J152621.70+432349.5 & 0.156 & 3.5 & 20.9 & 0.10 & 2499 & $35^{+13}_{-8}$ & $0.05\pm0.01$ & 0.04 & 0.03 \\
SDSS J151925.59+440755.8 & 0.238 & 3.8 & 39.8 & 0.16 & 2260 & $146^{+130}_{-47}$ & $0.10\pm0.03$ & 0.04 & 0.07 \\
SDSS J151526.86+523559.2 & 0.241 & 12.5 & 10.3 & 0.02 & 5470 & $49^{+22}_{-13}$ & $0.16\pm0.02$ & 0.11 & 0.11 \\
SDSS J140632.57+454004.8 & 0.241 & 5.2 & 22.2 & 0.07 & 3011 & $141^{+145}_{-47}$ & $0.24\pm0.07$ & 0.10 & 0.26 \\
SDSS J142034.56+452242.1 & 0.212 & 5.5 & 6.3 & 0.02 & 4108 & $153^{+159}_{-55}$ & $0.09\pm0.03$ & 0.04 & 0.04 \\
SDSS J153252.96+384330.5 & 0.134 & 1.2 & 10.4 & 0.16 & 1729 & $86^{+45}_{-23}$ & $0.07\pm0.01$ & 0.03 & 0.05 \\
SDSS J114832.32+580250.9 & 0.192 & 1.2 & 5.3 & 0.09 & 2032 & $65^{+43}_{-23}$ & $0.11\pm0.02$ & 0.08 & 0.05 \\
SDSS J144110.01+522557.0 & 0.339 & 11.5 & 20.4 & 0.03 & 4495 & $153^{+184}_{-57}$ & $0.19\pm0.06$ & 0.10 & 0.08 \\
SDSS J153800.30+461455.3 & 0.212 & 6.1 & 15.5 & 0.04 & 3523 & $149^{+165}_{-51}$ & $0.12\pm0.04$ & 0.06 & 0.06 \\
SDSS J133627.97+442917.7 & 0.137 & 1.3 & 9.7 & 0.14 & 1848 & $63^{+32}_{-17}$ & $0.10\pm0.02$ & 0.05 & 0.05 \\
SDSS J140829.93+414533.8 & 0.175 & 5.9 & 13.9 & 0.04 & 3543 & $34^{+16}_{-10}$ & $0.06\pm0.01$ & 0.06 & 0.04 \\
SDSS J102204.52+430809.6 & 0.315 & 4.3 & 27.5 & 0.10 & 2602 & $101^{+82}_{-35}$ & $0.07\pm0.02$ & 0.07 & 0.03 \\
\enddata
\tablecomments{Only the first 25 rows are presented here. The full table is available in CSV format.}
\tablenotetext{a}{The uncertainty on all $M_{\mathrm{BH}}$ values is a factor of 2 ($\pm{0.3}$ dex).}
\label{tab:les127}
\end{deluxetable*}

\begin{figure}[ht!]
\centering
\includegraphics[width=0.5\textwidth]{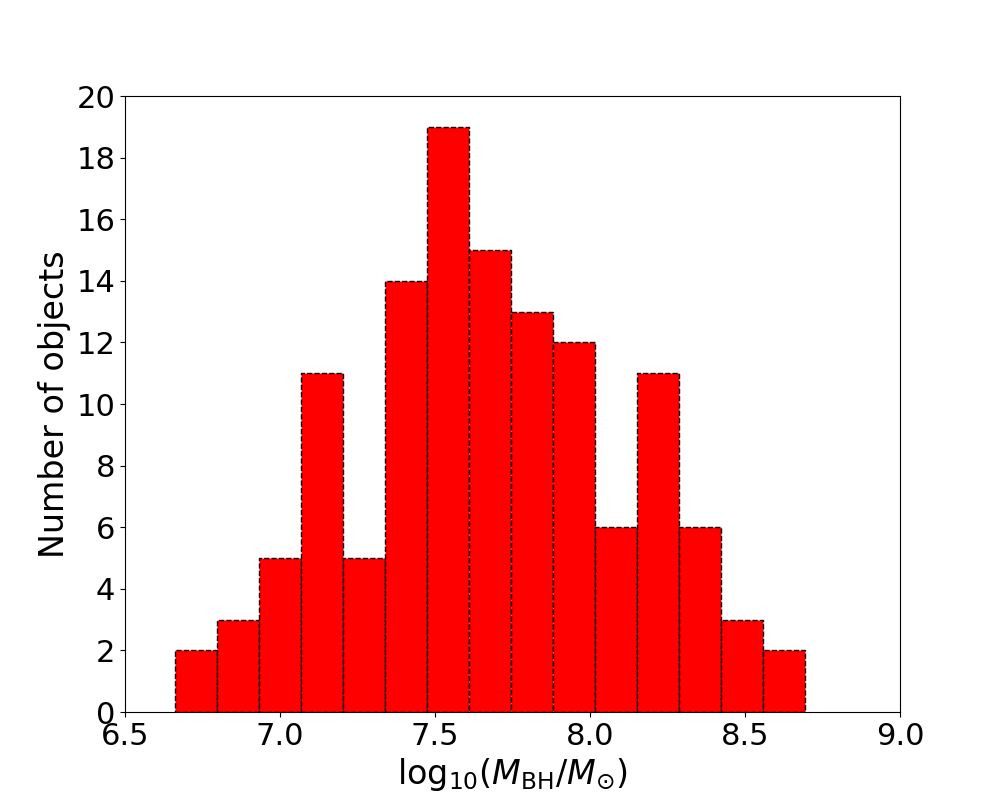}
\caption{Histogram of the black hole masses for the final sample of 127 AGNs. The sample is composed of less massive AGNs than the GLEAN AGNs inside the DR16Q catalog of \citet{WuShen2022}.}
\label{fig:density}
\end{figure}

\begin{figure}[ht!]
\centering
\includegraphics[width=0.5\textwidth]{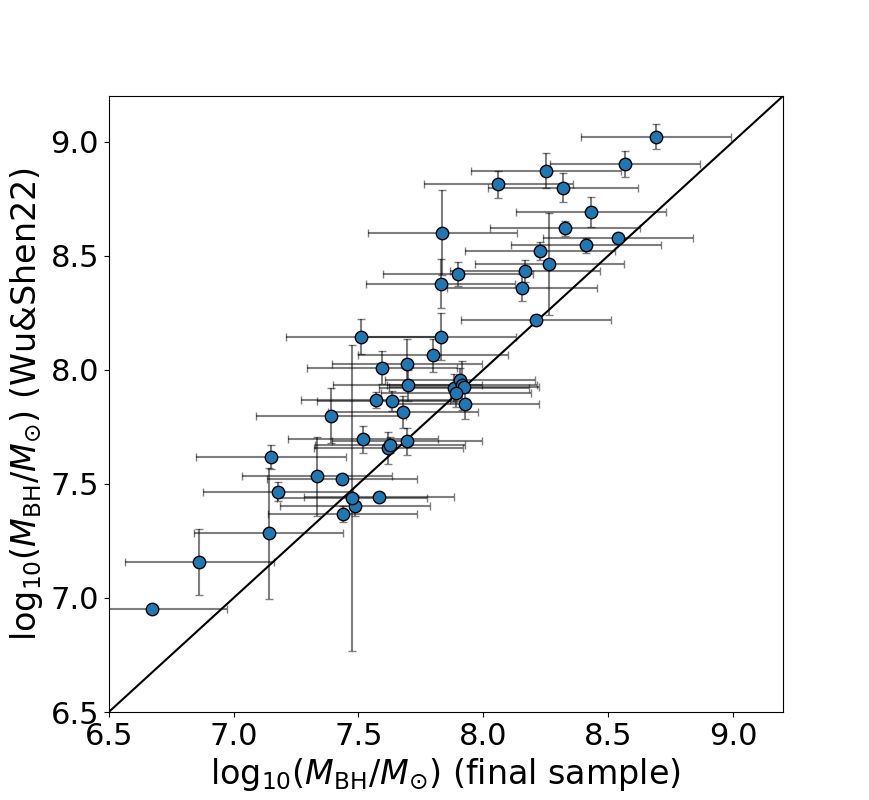}
\caption{Comparison of the black hole masses of 48 objects of our final sample present in the DR16Q catalog of \citet{WuShen2022}. The average absolute deviation of BH masses is 0.25 dex between the two measurements, and the deviation is of the same order of magnitude across the full range of masses.}
\label{fig:wushen vs us}
\end{figure}

Figure~\ref{fig:SDSS spectra} shows the SDSS spectra of four GLEAN AGNs, and the best multi-component Gaussian model associated with each. For the vast majority, SDSS spectra of GLEAN AGNs have broad H$\alpha$, but some do not have sufficient signal-to-noise ratio to measure black hole masses. Table~\ref{tab:les127} summarizes the parameters for each AGN in the final sample, while Figure~\ref{fig:density} shows the black hole mass distribution of the final sample of 127 AGNs. The masses peak at log$_{10}(M_{\mathrm{BH}}/M_{\odot}) = 7.5$ and vary between 6.5 and 9. They agree within the uncertainties of the values from \citet{WuShen2022} for the most part, as shown on Fig.~\ref{fig:wushen vs us}. For the 48 AGNs present in DR16Q out of 127, the average deviation from the \citet{WuShen2022} BH masses is 0.23 dex, which is within our systematic uncertainties of 0.3 dex. 14 AGNs are above 0.3 dex of deviation, with 0.76 dex being the largest difference. We believe this systematic difference in BH masses stems from the fact that we use the updated formula from \citep{Chilingarian2018} while \citet{WuShen2022} use the one from \citet{Greene2005} which has the original coefficients, and systematically gives slightly higher BH masses (about 0.1-0.15 dex of difference on average) for the same values of $L_{\mathrm{H\alpha}}$ and FWHM$_{\mathrm{H\alpha}}$. \citet{WuShen2022} also include the [\ion{S}{2}] doublet at 671.8 and 673.2 nm in their H$\alpha$ fitting, while we do not. However, the wavelength range used for the broad H$\alpha$ fitting does not overlap with [\ion{S}{2}], and so this difference is likely not significant.\\

\begin{figure*}[ht!]
\centering
\includegraphics[width=\textwidth]{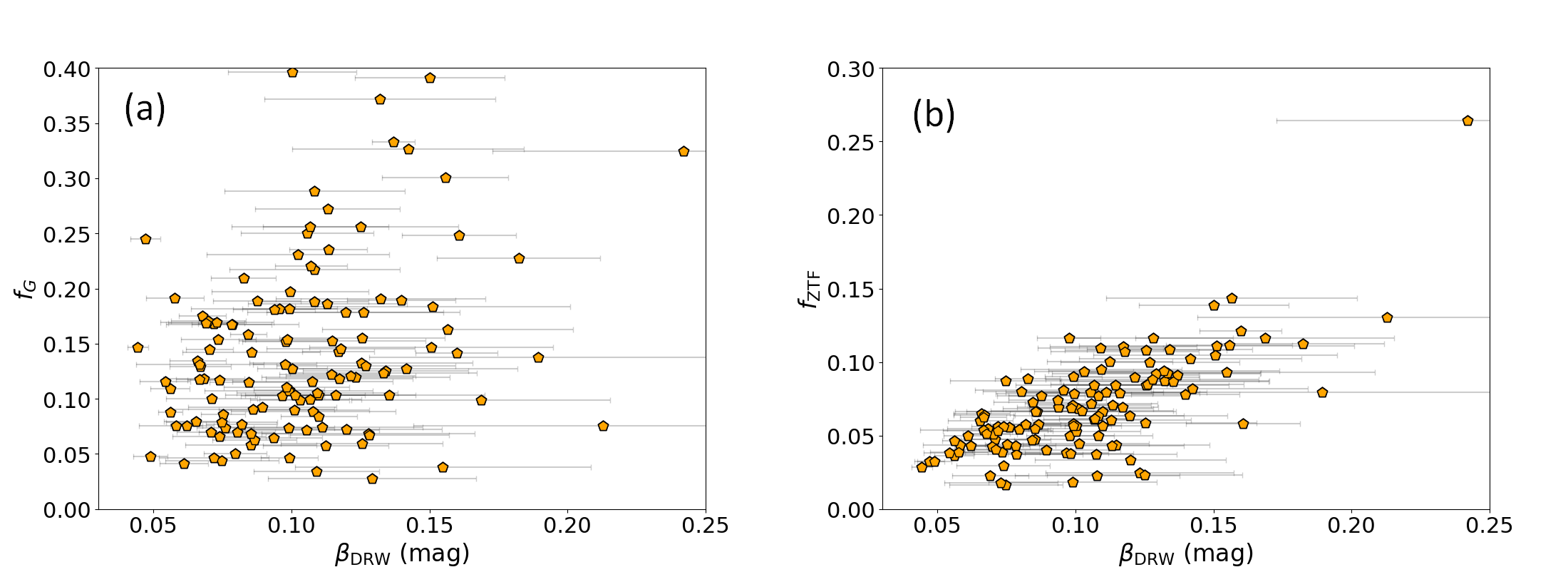}
\caption{DRW amplitude (calculated from ZTF light curves) versus fractional variability amplitude from (a) GLEAN, and from (b) ZTF, for the 127 GLEAN-SDSS-ZTF AGNs. Unlike panel (a) where the Spearman correlation coefficient $\rho$ between the two parameters is 0.237, they correlate well in panel (b), with $\rho = 0.668$.}
\label{fig:comparison frac var}
\end{figure*}

In Figure~\ref{fig:comparison frac var}, we compare the DRW amplitude calculated from ZTF to the fractional variability amplitude using GLEAN ($f_G$) and ZTF light curves ($f_{\mathrm{ZTF}}$), for the 127 AGNs of our final sample. To calculate $f_{\mathrm{ZTF}}$, we use the same formula that was used to get $f_G$ in the GLEAN catalog (Eq.~\ref{eq:f_G}). We check how each quantity evolves with $\beta_{\mathrm{DRW}}$. In Figure~\ref{fig:comparison frac var}a, the points are very dispersed and there is no clear relationship between $\beta_{\mathrm{DRW}}$ with $f_G$. The Spearman correlation coefficient is $\rho = 0.237$. However, in Figure~\ref{fig:comparison frac var}b, we have a strong correlation of $\beta_{\mathrm{DRW}}$ with $f_{\mathrm{ZTF}}$, with $\rho = 0.668$. This result therefore confirms the usefulness of the fractional variability amplitude as a model-independent measure of the amplitude of variability, but only when computed from high-quality light curves.\\

\begin{figure*}[ht!]
\centering
\includegraphics[width=\textwidth]{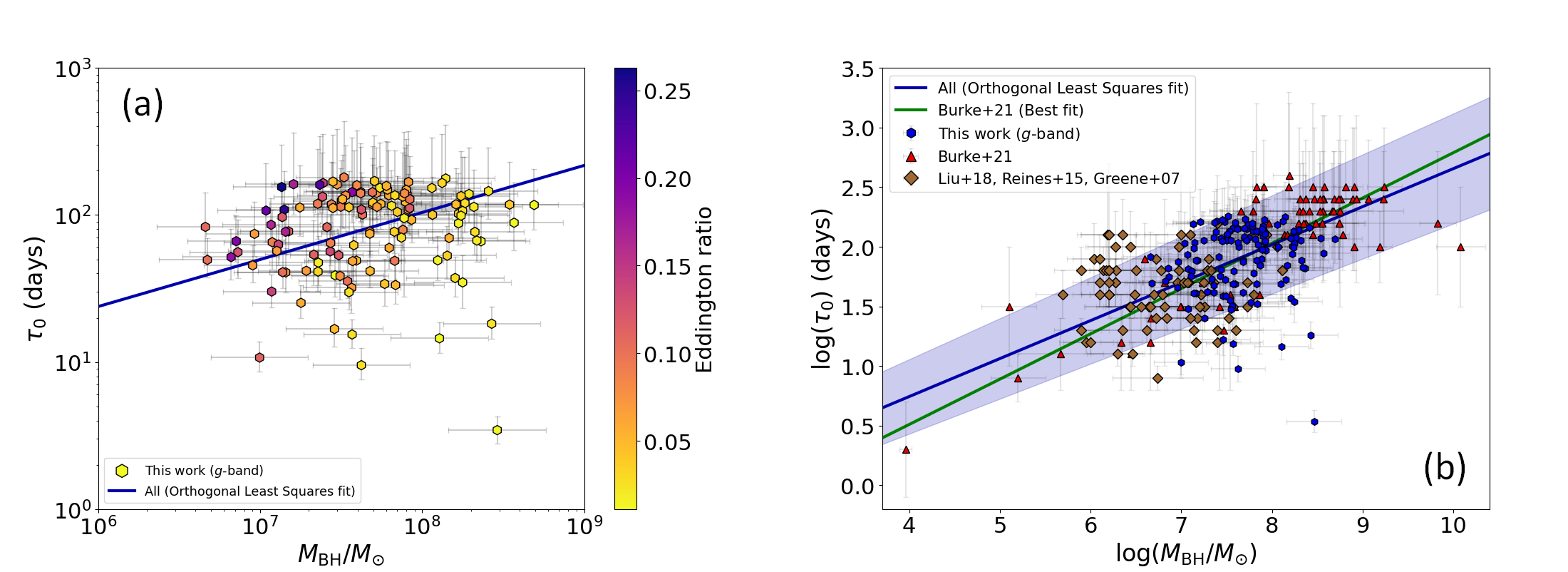}
\caption{Rest-frame timescale versus black hole mass for the 127 GLEAN AGNs. In (a), a colorbar shows the Eddington ratio of each AGN. The Eddington ratio of an AGN tends to increase when the $M_{\mathrm{BH}}$ of the AGN decreases in flux-limited samples. In (b), we add relevant samples from the literature used in \citet{Wang2023}. The blue background area shows the uncertainties in the slope and intercept value of the fit (blue line). The green line is the best fit from B21. The correlation between $\tau_0$ and $M_{\mathrm{BH}}$ is quite constrained, only with a few outliers, and the average timescale uncertainty of our data is comparable to the other datasets.}
\label{fig:edd+bces}
\end{figure*}

Figure~\ref{fig:edd+bces} presents the plot of rest-frame timescales and black hole masses. In Figure~\ref{fig:edd+bces}a, we calculated the Eddington ratio $L_{\mathrm{bol}}/L_{\mathrm{Edd}}$ of each one of the 127 AGNs. $L_{\mathrm{bol}}$ is obtained with the following approximation, assuming $L_{\mathrm{bol}} \simeq 9.8L_{\mathrm{5100A}}$, the continuum luminosity at 5100 $\mathrm{\AA}$ \citep{McLureDunlop2004,Greene2007}:
\begin{equation}
    L_{\mathrm{bol}} = 2.34 \times 10^{44} \left(\frac{L_{\mathrm{H\alpha}}}{10^{42}~\mathrm{erg~s^{-1}}} \right)^{0.86} \mathrm{erg~s^{-1}}~,
\end{equation}
and $L_{\mathrm{Edd}}$ is given by:
\begin{equation}
    L_{\mathrm{Edd}} = 1.26 \times 10^{38} \left(\frac{M_{\mathrm{BH}}}{M_{\odot}} \right) \mathrm{erg~s^{-1}}.
\end{equation}

We see that the Eddington ratio of an AGN tends to increase when the $M_{\mathrm{BH}}$ of the AGN decreases. This is expected, because lower-mass AGNs have to produce more light to be detectable above the flux limit of Gaia and SDSS. In Figure~\ref{fig:edd+bces}b, we add the 67 AGNs from B21 and the 79 dwarf galaxy AGNs from \citet{Liu2018}, \citet{Reines2015}, and \citet{Greene2007}, alongside our data. Before plotting, we convert from the observed-frame DRW timescale to a rest-frame DRW timescale $\tau_0 = \tau_{\mathrm{DRW}}/(1+z)$, which accounts for cosmological time dilation. The correlation between $\tau_0$ and $M_{\mathrm{BH}}$ is quite constrained, with only a few outliers, and the average timescale uncertainty of our data is comparable to the other datasets. In Figure~\ref{fig:edd+bces}, we added the best fit model of B21 which identified a correlation between the rest-frame DRW timescale and the black hole mass:
\begin{equation}\label{eq:burke correlation}
    \tau_0 = 107^{+11}_{-12}\left( \frac{M_{\mathrm{BH}}}{10^8 M_{\odot}}\right)^{0.38^{+0.05}_{-0.04}} \mathrm{days}.
\end{equation}
Inverting the relation, we obtain:
\begin{equation}
    M_{\mathrm{BH}} = 10^{6.61^{+0.11}_{-0.13}}\left( \frac{\tau_0}{32~\mathrm{days}}\right)^{2.63 \pm 0.31} M_{\odot}.
\end{equation}

We determine our own best-fit line from the entire set of data points (273 AGNs). We use the Orthogonal Least Squares (OLS) method \citep{Akritas1996} with the $bces$ package\footnote{\href{https://pypi.org/project/bces/}{https://pypi.org/project/bces/}} \citep{BCES2012}, which has the advantage of taking into account error bars in both $x$ and $y$ variables. The intrinsic scatter of the correlation is well-represented by the dark blue area, which indicates the uncertainties in the slope and intercept values. Writing the resulting formula for the best-fit line in the same format as Equation~\ref{eq:burke correlation}, we find
\begin{equation}\label{eq:our correlation}
    \tau_0 = 104^{+66}_{-40}\left( \frac{M_{\mathrm{BH}}}{10^8 M_{\odot}}\right)^{0.32 \pm 0.03} \mathrm{days}.
\end{equation}
Once again inverting the relation, we obtain:
\begin{equation}\label{eq:our correlation bis}
    M_{\mathrm{BH}} = 10^{6.28 \pm 0.54}\left( \frac{\tau_0}{32~\mathrm{days}}\right)^{3.16 \pm 0.25} M_{\odot}.
\end{equation}
Our intercept value of 104 days is very close to the B21 value of 107 days. The larger scatter comes directly from the use of the OLS method, and the larger number of data points. Our slope ($0.32 \pm 0.03$) indicates a slightly weaker dependence of $\tau_0$ on $M_{\mathrm{BH}}$, but is consistent with the B21 value of $0.38^{+0.05}_{-0.04}$ given the uncertainties. The scatter would probably decrease a little bit if we were using the Maximum Likelihood Estimator (MLE) based on a Gaussian mixture model, described in \citet{Kelly2007}. The Figure 5 of \citet{Kelly2007} shows that the standard deviation of the intrinsic scatter is slightly lower for the MLE method than the OLS method (in the sense that the peak of the MLE distribution is at a lower value than the OLS).\\

\section{Discussion} \label{sec:discussion}

The ZTF-measured variability timescales of GLEAN AGNs are consistent with the $\tau_0$-$M_{\mathrm{BH}}$ correlation identified in previous works, and we add 127 additional AGNs with $M_{\mathrm{BH}}$ and $\tau_0$ values to the works of \citet{Wang2023} and B21, raising the total sample size from 146 to 273 AGNs. This improves the accuracy of the fit parameters. For the past 15 years, many authors \citep{Kelly2009,Macleod2010,Kozlowski2016,Simm2016,Guo2017,Burke2021,Suberlak2021,Arevalo2024,Su2024} have proposed a broad range of slope values for the $\tau_0$-$M_{\mathrm{BH}}$ correlation (from $\sim 0$ to $0.8$), summarized by \citet{Su2024}. There does not seem to be any consensus, possibly due to the intrinsic scatter of the relationship itself. Our slope and intercept are closest to those of B21 ($0.38^{+0.05}_{-0.04}$) and \citet{Kozlowski2016} ($0.38\pm0.15$), but the main limit of B21's work is the sparse number of data points on the low $M_{\mathrm{BH}}$/short $\tau_0$ part of the correlation. As we do not find low mass black holes in our final sample, we cannot independently comment on whether this relationship extends to the lower mass population of black holes.\\

Removing sources above a certain variability timescale because of the baseline condition potentially biases the sample mean towards lower timescales, particularly for higher black hole masses. Indeed, if we omit the baseline condition, the average timescale would increase, by including AGNs with longer estimated timescales at the same mass. However, the reliability of the longer timescale measurements decreases significantly, which highlights the importance of setting this condition. Similarly, even if the upper limit on the variability timescale is dependent on the baseline and the dataset, we still have evidence in Figure~\ref{fig:edd+bces}b that the variability timescale correlates with the black hole mass, with three different datasets. As mentioned in B21, this caveat does not affect the existence of the correlation, because they found the same correlation in a larger and lower quality sample (with a black hole mass slope of $\sim 0.5$).\\

The fractional variability, $f_G$, from GLEAN light curves does not correlate with $\beta_{\mathrm{DRW}}$, but when we calculate the fractional variability with the ZTF light curves ($f_{\mathrm{ZTF}}$), we find that $f_{\rm ZTF}$ correlates quite well with $\beta_{\mathrm{DRW}}$. This means that the $f_G$ values from the GLEAN catalog are not indicative of the actual amplitude of variability of the AGN. We believe the discrepancy comes from the cadence of observations, which is relatively sparse for GLEAN light curves. On the other hand, ZTF light curves have many more data points that allow a better traceback of amplitude variations, hence, of the fractional variability amplitude. $f_{\mathrm{ZTF}}$ values are a good proxy for $\beta_{\mathrm{DRW}}$, are model-independent and straightforward to compute. Therefore, we recommend the use of the fractional variability amplitude parameter as a proxy of the variability amplitude for sufficiently well-sampled light curves.\\

Figure~\ref{fig:ibandz} shows that while GLEAN's purity is very high, there are some contaminants according to SDSS classifications. Perhaps the most intriguing objects are probably the 0.12\% classified as stars in SDSS but as AGNs in Gaia. Because they are in GLEAN, these star candidates have no detectable proper motion ($< 1~\mathrm{mas~yr^{-1}}$), they lack reliable parallax measurements (some of them are negative), and they were variable in the optical between 2014 and 2017. These are not the characteristics of typical stars. A thorough inspection of the SDSS spectra and photometry of these objects could perhaps unveil some true AGNs that were unidentified as such by SDSS due to a wrong spectral fit. This will be the subject of future work.\\

\section{Conclusions}
\label{sec:conclusion}

We characterize the GLEAN catalog of $872~228$ AGNs selected on variability and observed by SDSS, using SDSS classifications and the quasar properties catalog from \citet{WuShen2022}. We also better constrain the correlation between the central black hole mass measured from broad H$\alpha$ in single-epoch SDSS spectra and the variability timescale from high-quality ZTF light curves with the addition of 127 GLEAN AGNs to previous samples, for a total number of 273 AGNs.
\begin{itemize}
  \item Combining our final curated sample of 127 AGNs with the works of \citet{Burke2021} and \citet{Wang2023}, we confirm the correlation between the rest-frame DRW timescale of variability and the black hole mass of the AGN. We perform our own linear fit to the combined distribution of 273 AGNs. We find a slope value of $0.32 \pm 0.03$, consistent with the slope of \citet{Burke2021} ($0.38^{+0.05}_{-0.04}$):
  \begin{equation}
    \tau_0 = 104^{+66}_{-40}\left( \frac{M_{\mathrm{BH}}}{10^8 M_{\odot}}\right)^{0.32 \pm 0.03} \mathrm{days}.
  \end{equation}
  \item We do not find any low-mass black holes in our final sample. According to the \citet{WuShen2022} catalog, the great majority of the AGNs in GLEAN with SDSS black-hole mass measurements have a central black hole mass between $10^8$ and $10^{9.5} M_\odot$.
  \item We do not recommend using GLEAN light curves for stochastic modeling, specifically with a damped random walk (DRW) model. The light curves are not sampled well-enough to allow a reliable estimation of the DRW timescale outside of the range of 50 to 100 days.
  \item Compared to GLEAN, ZTF light curves with $n_{\rm obs} > 1000$ allow a more accurate estimation of the DRW timescale and amplitude, due to their high sampling.
  \item The fractional variability amplitudes from GLEAN light curves do not correlate with the DRW amplitudes. However, DRW amplitudes do correlate well with the fractional variability amplitude using ZTF light curves, $f_{\mathrm{ZTF}}$. The fractional variability is therefore a good proxy for the amplitude of variability in AGNs, but only with ZTF-quality light curves.
\end{itemize}

\begin{acknowledgments}

We would like to acknowledge that the University of Western Ontario, where the research presented in this article has been produced, is located on the traditional lands of the Anishinaabek, Haudenosaunee, L$\mathrm{\bar{u}}$naap$\mathrm{\acute{e}}$ewak and Chonnonton Nations, on lands connected with the London Township and Sombra Treaties of 1796 and the Dish with One Spoon Covenant Wampum. These peoples have been the original caretakers of this land, and we should give them the utmost respect in that regard. We also acknowledge historical and ongoing injustices that Indigenous Peoples (First Nations, M$\mathrm{\acute{e}}$tis and Inuit) endure in Canada, and we recognize the long way to go towards reconciliation and decolonization.\\

We thank the anonymous referee for their helpful comments on the article. AH, PB and SG acknowledge Discovery Grant support from the Natural Sciences and Engineering Research Council of Canada (NSERC). We acknowledge the use of the following software and Python packages: TOPCAT \citep{Taylor2005}, astropy \citep{Astropy2018} and \textit{taufit} \citep{Burke2021}.

This work has made use of data from the European Space Agency (ESA) mission \href{https://www.cosmos.esa.int/gaia}{Gaia}, processed by the \href{https://www.cosmos.esa.int/web/gaia/dpac/consortium}{Gaia Data Processing and Analysis Consortium} (DPAC). Funding for the DPAC has been provided by national institutions, in particular the institutions participating in the Gaia Multilateral Agreement. The Gaia archive website is \href{https://archives.esac.esa.int/gaia}{https://archives.esac.esa.int/gaia}.

This work has also used data from the Sloan Digital Sky Survey. Funding for the Sloan Digital Sky Survey IV has been provided by the Alfred P. Sloan Foundation, the U.S. Department of Energy Office of Science, and the Participating Institutions. SDSS-IV acknowledges support and resources from the Center for High Performance Computing at the University of Utah. The SDSS website is \href{www.sdss4.org}{www.sdss4.org}.

ZTF is supported by the National Science Foundation under Grants No. AST-1440341 and AST-2034437 and a collaboration including current partners Caltech, IPAC, the Oskar Klein Center at Stockholm University, the University of Maryland, University of California, Berkeley, the University of Wisconsin at Milwaukee, University of Warwick, Ruhr University, Cornell University, Northwestern University and Drexel University.

\end{acknowledgments}

\clearpage

\bibliography{citations}{}
\bibliographystyle{aasjournal}

\end{document}